\documentclass[aps,pra,reprint,superscriptaddress,singlecolumn,amsmath,amssymb]{revtex4-1}

\usepackage{graphicx}
\usepackage{dcolumn}
\usepackage{bm}
\usepackage{xcolor}
\usepackage{braket}
\usepackage{comment}
\usepackage{amsfonts} 
\usepackage{amsmath}
\usepackage{amssymb}
\usepackage{bbm}
\usepackage{coffeestains}
\usepackage{graphicx}
\usepackage{ragged2e}
\usepackage[colorlinks=true, linkcolor=blue, citecolor=blue, urlcolor=blue]{hyperref}

\begin{document}

\title{Perturbative approach to the first law of quantum thermodynamics}

\author{Mario Reis}
\email{marioreis@id.uff.br}
\affiliation{Instituto de Física, Universidade Federal Fluminense, Av. Gal. Milton Tavares de Souza s/n, 24210-346, Niteroi-RJ, Brazil}
\author{Maron F. Anka}
\email{maron.anka@fbter.org.br}
\affiliation{QuIIN - Quantum Industrial Innovation, EMBRAPII CIMATEC Competence Center in Quantum Technologies, SENAI CIMATEC, Av. Orlando Gomes 1845, Salvador, BA, Brazil, CEP 41650-010.}
\author{Vinicius Gomes de Paula}
\email{viniciusgomespaula@id.uff.br}
\affiliation{Escola de Engenharia de Petrópolis, Universidade Federal Fluminense, Rua Domingos Silvério, s/n, 25650-050, Quitandinha - RJ, Brazil}
\author{Clebson Cruz}
\email{clebson.cruz@ufob.edu.br}
\affiliation{Centro das Ciências Exatas e das Tecnologias, Universidade Federal do Oeste da Bahia, Rua Bertioga 892, Morada Nobre, 47810-059}

\date{\today}

\begin{abstract}
In quantum thermodynamics, the decomposition of energy exchanges into heat and work remains an open problem beyond weak-coupling and slow-driving regimes. Recent formulations have shown that quantum coherence introduces additional energy contributions whose thermodynamic interpretation is still under debate, raising fundamental questions about the structure of the quantum first law. In this work, we investigate this problem through a time-dependent perturbative framework applied to the first law of quantum thermodynamics. By expanding the thermodynamic quantities up to second order, we derive explicit perturbative corrections for work, heat, and coherence contributions. Our results show that the coherence term can be consistently decomposed into coherent heat and coherent work, demonstrating that quantum coherence does not require the introduction of an independent energetic contribution beyond heat and work. The formalism resolves inconsistencies associated with previous formulations of the quantum first law, including the interpretation of coherence contributions and their connection with entropy fluxes. At second order, the perturbative corrections become directly connected to transition rates governed by Fermi’s golden rule, establishing a bridge between microscopic quantum transitions and macroscopic thermodynamic quantities. These results provide a physically transparent framework to investigate coherence-driven thermodynamic processes and offer new perspectives for the analysis of driven quantum systems and nonequilibrium quantum technologies.

\end{abstract}

\maketitle

\section{Introduction}

Quantum thermodynamics is an exciting and interdisciplinary field that merges quantum mechanics, non-equilibrium statistical physics, classical stochastic thermodynamics, and information theory in order to understand energy exchange processes in quantum systems \cite{binder2018thermodynamics,deffner2019quantum,strasberg2022quantum}. There are numerous efforts to explore several quantum features in thermodynamic processes, for instance, quantum measurements \cite{yi2017single,ding2018measurement,jordan2020quantum,elouard2017role}, quantum coherence \cite{streltsov2017colloquium, santos2019role,francica2019role,francica2020quantum,gour2022role} and quantum correlations \cite{touil2021ergotropy,francica2022quantum,cruz2022quantum,beny2018energy,cruz2023quantum,moroder2024thermodynamics}, among others \cite{rossnagel2014nanoscale,myers2020bosons,ciccarello2022quantum}. This field emerges as a natural framework to deal with the modern development of current technology, originating from the second quantum revolution \cite{jaeger2018second,atzori2019second,deutsch2020harnessing,auffeves2022quantum}. The constant scaling-down in size of new technologies brings out the necessity of dealing with energy exchanges with not only thermal but also quantum fluctuations, which are inherently probabilistic by nature. This problem complicates the practical application of quantum technology due to the need to maintain quantum systems in specific conditions, such as extremely low temperatures \cite{patra2017cryo,krinner2019engineering}. Nevertheless, significant advances have been made in the field, including the development of quantum versions of classical thermal machines \cite{quan2007quantum,quan2009quantum,kosloff2014quantum,ghosh2018thermodynamic}, the investigation of the role of quantum coherence as a thermodynamic resource \cite{scully2003extracting,uzdin2015equivalence,lostaglio2015description,korzekwa2016extraction,kammerlander2016coherence}, and the establishment of deep connections between quantum information theory and thermodynamics \cite{leff2002maxwell,sagawa2012thermodynamics,parrondo2015thermodynamics}. Further progress has been achieved through the derivation of quantum fluctuation theorems \cite{campisi2009fluctuation,campisi2011colloquium,aaberg2018fully,micadei2020quantum}, the formulation of a general framework for entropy production \cite{landi2021irreversible}, and the remarkable experimental demonstration that non-classical correlations can drive heat to flow against its natural direction — from cold to hot — without any external work input \cite{micadei2019reversing}. For a clear review on recent progress in theoretical and experimental quantum thermodynamics, we refer the reader to Refs. \cite{myers2022quantum,campbell2026roadmap}.

Despite significant advances, fundamental concepts remain open problems in quantum thermodynamics. While the basic notions of work and heat are well established and consolidated in classical and even stochastic thermodynamics \cite{sekimoto2010stochastic,peliti2021stochastic,shiraishi2023introduction}, they remain ill-defined quantities in the quantum regime. This difficulty stems from the very nature of work and heat: unlike internal energy, they are not state functions but rather process functions — path-dependent quantities that characterize energy exchange along a thermodynamic process, not the state of the system itself. As a consequence, work and heat cannot be associated with Hermitian operators \cite{talkner2007fluctuation}, since such operators would imply well-defined, path-independent observables. Furthermore, in the quantum domain, the notion of a trajectory is lost (in the classical sense), making it unclear how to unambiguously distinguish the energy exchanged in the form of work from that exchanged as heat in a general quantum thermodynamic process \cite{binder2018thermodynamics}. Consequently, a significant portion of the literature on quantum thermal devices involving thermodynamic processes explores scenarios where the production of work and heat occurs separately, e.g., the quantum Otto cycle \cite{kosloff2017quantum,pena2020otto,solfanelli2020nonadiabatic,anka2021measurement,piccitto2022ising,cherubim2022nonadiabatic}. A notable exception arises in the weak-coupling Markovian regime, where the separation of timescales between system and bath, combined with the secular approximation, yields a natural and unambiguous decomposition of energy exchange into work and heat, as derived in Alicki's pioneering paper \cite{alicki1979quantum} — yet even this well-established framework breaks down for fast driving, where the instantaneous form of the first law is only recovered in the cyclic operations limit \cite{binder2018thermodynamics}.

Most recently, extensive efforts have been put forward to provide a general framework of the first law of quantum thermodynamics with an unambiguous definition of work and heat beyond the weak-coupling, Markovian dynamics with a slow driving regime \cite{alicki1979quantum}. These redefinitions were independently proposed by two groups, grounded in the following bases \cite{alipour2022entropy,ahmadi2023contribution}: the changes originated from the Hamiltonian, and the variation of the eigenstates are treated as work contributions, and the heat is associated with the entropy variation. Nevertheless, the adequacy of such redefinitions was questioned through two counterexamples presented in Ref. \cite{botosso2021comment}. In the first example, involving two interacting systems, although no consensus framework currently exists for a proper thermodynamic description, there is no energy being exchanged at the same time that the entropy of one of the subsystems changes. In the second, the redefinitions fail to reproduce the standard expected behavior of a qubit interacting with a zero-temperature Markovian reservoir, in contrast with the expected description provided by Alicki’s original framework. One year later\footnote{Although Refs. \cite{alipour2022entropy,ahmadi2023contribution} were only published in 2022 and 2023, respectively, they were first introduced in 2019, whereas Bertúlio’s work dates to 2020.}, a third framework was introduced in which the rotation of the eigenvectors is associated with the contribution arising from quantum coherence \cite{de2020unraveling}. Within this approach, the first law is no longer composed solely of heat and work contributions but also includes an additional term — referred to as coherent energy — associated with the quantum coherence present in the system.

In this work, we further explore the third approach. First, we demonstrate the equivalence between the frameworks of Alipour \textit{et al.} \cite{alipour2022entropy} and Ahmadi, Salimi, and Khorashad \cite{ahmadi2023contribution} and that of Bertúlio \cite{de2020unraveling}: while the former associates the eigenvector contribution with work, the latter attributes it to quantum coherence, differing only in the interpretation and physical origin assigned to this quantity. 
{In the following, we then employ perturbation theory to derive the work, heat, and coherence contributions up to second order. In particular, it is shown that the coherent term can be decomposed into distinct work and heat contributions, thereby shedding light on the role of quantum coherence in quantum thermodynamics and demonstrating that the first law of thermodynamics in the quantum regime is still comprised of work and heat boosted by a genuine quantum feature.} 

\section{Background}

A canonical setting in quantum thermodynamics consists of an open quantum system, composed by a subsystem of interest $\hat{\rho}$, e.g., a qubit, coupled to one or more environments $\hat{\rho}_{E_1,...E_N}$. In the system-environment weak coupling limit, the dynamics of the system $\hat{\rho}$ is well described by the so-called Gorini-Kossakowski-Sudarshan-Lindblad master equation (or just Lindblad master equation) \cite{breuer2002theory,lidar2019lecture}
\begin{equation}
\frac{d\hat{\rho}(t)}{dt} = - \dfrac{i}{\hbar} [\hat{H}(t),\hat{\rho}(t)] + \mathcal{L}[\hat{\rho}(t)],
\end{equation}
where $\hbar$ is the reduced Planck constant, $\hat{H}(t)$ is the Hamiltonian of the system, the first term is the von Neumann evolution equation associated with the reversible dynamics and $\mathcal{L}[\hat{\rho}(t)] = \sum_m \mathcal{L}_m[\hat{\rho}(t)]$ is the dissipative part of the dynamics due to the interaction with the environments, giving rise to decoherence effects. The decomposition of the dynamics into a coherent and a dissipative contribution provides a natural starting point for defining thermodynamic quantities at the quantum level. Following Alicki's seminal work \cite{alicki1979quantum}, the energy conservation law for the system under a open dynamics reads
\begin{equation}
\dfrac{dU(t)}{dt} = \mathcal{P}(t) + \mathcal{J}(t),
\end{equation}
where
\begin{equation}
\begin{split}
U(t) &= \text{tr}[\hat{\rho}(t) \hat{H}(t)], \\
\mathcal{P}(t) &= \text{tr}\Big[\hat{\rho}(t) \frac{d\hat{H}(t)}{dt}\Big], \\
\mathcal{J}(t)&= \text{tr}\Big[\frac{d\hat{\rho}(t)}{dt} \hat{H}(t)\Big]
\end{split}
\end{equation}
are the internal energy of the system of interest, the power injected by the external driving, and the heat current exchanged with the environments, respectively. The average work and heat over a time interval $[0, t]$ are then defined as
\begin{align}
\langle W \rangle_t \equiv W(t) &:= \int_0^t dt' \text{tr}\Big[\hat{\rho}(t') \frac{d\hat{H}(t')}{dt'}\Big], \label{alickiwork}\\
\langle Q \rangle_t \equiv Q(t)&:= \int_0^t dt'  \text{tr}\Big[\frac{d\hat{\rho}(t')}{dt'} \hat{H}(t')\Big]. \label{alickiheat}
\end{align}
In the $\hat{H}(t)$ energy eigenbasis $\{\ket{n(t)}\}$, the above equations reduce to
\begin{align}
W(t) &= \int_0^t dt' \sum_n \rho_n(t') \dfrac{dE_n(t')}{dt}, \label{alickiwork2}\\
Q(t)&= \int_0^t dt' \sum_n \dfrac{d\rho_n(t')}{dt} E_n(t'), \label{alickiheat2}
\end{align}
where $E_n(t) = \bra{n(t)}\hat{H}(t)\ket{n(t)}$ and $\rho_n(t) = \bra{n(t)}\hat{\rho}(t)\ket{n(t)}$. The first law of quantum thermodynamics can simply be written as $\Delta U = U(t) - U(0) = \sum_n [p_n(t) E_n(t) - p_n(0) E_n(0)] = W(t) + Q(t)$.

The more recent entropy-based formulation of the first law was motivated by (i) extending general definitions of work and heat beyond the weak-coupling Markovian dynamics with a constant or slow driving Hamiltonian, (ii) the need to account for work exchanged through time-independent Hamiltonians, which Alicki's framework assigns identically zero, and (iii) the recognition that, in the presence of quantum coherence, the energy-basis populations no longer coincide with the eigenvalues of the density matrix operator, which solely determine the von Neumann entropy. As a consequence, part of the energy conventionally assigned to heat — whether viewed from the dissipative part of the dynamical equation or from population changes in the energy eigenbasis — does not contribute to entropy change and should instead be identified as work \cite{alipour2022entropy,ahmadi2023contribution}. In this framework, the infinitesimal heat and work are reformulated in terms of the spectral decomposition of the density matrix, $\hat{\rho}(t) = \sum_k \rho_k(t) \ket{k(t)}\bra{k(t)}$,ensuring that heat is exclusively tied to entropy-changing contributions
\begin{align}
\delta W(t) &= \sum_k \rho_k(t) d(\bra{k(t)}\hat{H}(t)\ket{k(t)}), \label{entropybasedw}\\
\delta  Q(t)&= \sum_k d\rho_k(t)  \bra{k(t)}\hat{H}(t)\ket{k(t)}, \label{entropybasedq}
\end{align}
where $\rho_k(t)$ and $\{\ket{k(t)}\}$ are the eigenvalues and eigenvectors of the density matrix, and $\delta$ denotes path-dependent quantities. Crucially, the heat is now governed by changes in the eigenvalues of $\hat{\rho}(t)$, which directly determine the von Neumann entropy, while the work captures both external driving and the entropy-preserving rotation of the state's eigenvectors.

In the third formalism introduced by Bertúlio \cite{de2020unraveling}, the quantum definitions of work and heat are derived from their classical counterparts. An inconsistency was demonstrated between two derivations of the first law: one based on the classical definition of work — computed in the energy basis $\{\ket{n}\}$ — and another based on the classical definition of heat — computed in the density matrix basis $\{\ket{k}\}$, where the time dependence of the eigenvectors has been omitted for simplicity. Despite the similarity between this and previous frameworks, an important distinction must be highlighted. Whereas the energy contribution arising from the derivative with respect to the eigenvectors in Eq.~\eqref{entropybasedw} is associated with work, Bertúlio directly links changes in the eigenvectors to quantum coherence, arguing that a consistent first law of thermodynamics for quantum systems must encompass a third energy contribution beyond work and heat — a path-dependent, coherence-based quantity. Thus, in this formulation, the first law becomes
\begin{equation}
\Delta U = W(t) + Q(t) + C(t),
\end{equation}
where
\begin{equation}\label{work}
W(t)=
\int_{0}^{t} dt' \sum_{n,k}
      \rho_{k}(t')\,\bigl|c_{nk}(t')\bigr|^{2}\,
      \frac{dE_{n}(t')}{dt'},
\end{equation}

\begin{equation}\label{heat}
Q(t)=
\int_{0}^{t} dt' \sum_{n,k}
      E_{n}(t')\,\bigl|c_{nk}(t')\bigr|^{2}\,
      \frac{d\rho_{k}(t')}{dt'},
\end{equation}

\begin{align}\label{coherence}
C(t) &=
\int_{0}^{t} dt' \sum_{n,k}
      E_{n}(t')\,\rho_{k}(t')\,
      \frac{d}{dt'}\!\bigl|c_{nk}(t')\bigr|^{2},
\end{align}
with $C(t)$ standing for the quantity identified as quantum coherence, and $c_{nk}(t) = \braket{n|k}$ being the ``transition coefficients" between the $\{\ket{n}\}$ and $\{\ket{k}\}$ basis. It is worth noticing that the coherence defined by Eq.~\eqref{coherence} has no classical analogue, being a purely quantum contribution for the energy exchange of the system.

Before moving on to the next section, we briefly discuss the implications of such approaches. The equivalence between Eq.~\eqref{entropybasedw} and Eqs.~\eqref{work} and \eqref{coherence} is directly obtained by substituting the explicit form of the Hamiltonian in the energy basis, $\hat{H}(t) = \sum_n E_n(t) \ket{n}\bra{n}$, in Eq.~\eqref{entropybasedw}
\begin{equation}
\begin{split}
\delta W(t) &= \sum_k \rho_k(t) d(\bra{k}\hat{H}(t)\ket{k}) \\
&= \sum_k \rho_k(t) d\Big(\bra{k}\Big(\sum_n E_n(t) \ket{n}\bra{n}\Big)\ket{k}\Big) \\
&= \sum_k \rho_k(t) d\Big(\sum_n E_n(t) |c_{nk}(t)|^2\Big) \\
&= \sum_{n,k} (\rho_k(t) |c_{nk}(t)|^2  d E_n(t) + \rho_k(t) E_n(t) d|c_{nk}(t)|^2).
\end{split}
\label{equivalencework}
\end{equation}
One can note that the quantum work from Refs. \cite{alipour2022entropy,ahmadi2023contribution} can be split into the quantum work and the coherence term from Ref. \cite{de2020unraveling}, implying that the difference between both frameworks arises from the interpretation of the second term of the last line in Eq.~\eqref{equivalencework}. Applying the same method to Eq.~\eqref{entropybasedq}, we obtain
\begin{equation}
\begin{split}
\delta Q(t) &= \sum_k d\rho_k(t)  \bra{k(t)}\hat{H}(t)\ket{k(t)} \\
&= \sum_k d\rho_k(t) \bra{k}\Big(\sum_n E_n(t) \ket{n}\bra{n}\Big)\ket{k} \\
&= \sum_{n,k} E_n(t) |c_{nk}(t)|^2 d\rho_k(t),
\end{split}
\label{equivalenceheat}
\end{equation}
which is the exact same result of Eq.~\eqref{heat}. While these results hint that the quantum coherence must be interpreted either as work or as an independent third energy contribution, Bertúlio argues that quantum coherence is an exclusive contribution to the quantum  version of the first law, independent from the work and heat quantities derived from their classical counterparts, Eqs.~\eqref{work} and \eqref{heat}. 

Another issue arises when we resort to the second law of thermodynamics, which can be recast as
\begin{equation}
\Pi = \dfrac{dS}{dt} + \phi \geq 0,
\label{secondlaw}
\end{equation}
where $\Pi$ is the entropy production, $dS/dt$ is the entropy rate, with $S$ being the von Neumann entropy, and $\phi$ is the entropy flux, which can be defined as \cite{santos2019role}
\begin{equation}
\phi = - \dfrac{\dot{Q}(t)}{T} = -\dfrac{1}{T} \sum_n E_n \dfrac{d\rho_n(t)}{dt},
\end{equation}
where $\dot{Q}(t) = dQ(t)/dt$, which is deeply related to Alicki's formulation of heat, Eq.~(\ref{alickiheat2}). Using the Ref. \cite{de2020unraveling} framework, where $\rho_n(t) = \sum_k \rho_k(t) |c_{nk}(t)|^2$, we obtain
\begin{equation}
\dfrac{d\rho_n(t)}{dt} = \sum_k  |c_{nk}(t)|^2 \dfrac{d\rho_k(t)}{dt} + \sum_k \rho_k(t) \dfrac{d|c_{nk}(t)|^2}{dt}.
\end{equation}
Hence, we have
\begin{align}
\begin{split}
\dot{Q}(t) &= \sum_n E_n(t) \dfrac{d\rho_n(t)}{dt}  \\&= \sum_{n,k}  E_n(t) |c_{nk}(t)|^2 \dfrac{d\rho_k(t)}{dt} + \sum_{n,k} E_n(t) \rho_k(t) \dfrac{d|c_{nk}(t)|^2}{dt} \\
&= \dot{Q}_{\text{pop}}(t) + \dot{Q}_{\text{coh}}(t),
\end{split}
\label{heatcurrent}
\end{align}
where each contribution allows us to define
\begin{equation}
\phi = \phi_{\text{pop}} + \phi_{coh} = -\dfrac{\dot{Q}_{\text{pop}}(t)}{T}  -\dfrac{\dot{Q}_{\text{coh}}(t)}{T},
\label{fluxes}
\end{equation}
where $\phi_{\text{pop}}$ and $\phi_{coh}$ are the entropy fluxes associated with the population and coherence changes, respectively. Notice that, while the entropy-based first law implies that the quantum coherence is associated with work, the second law based on Alicki's definition of heat attributes it to an entropy flux. In this sense, the formalism from Refs. \cite{alipour2022entropy,ahmadi2023contribution,de2020unraveling} lacks any direct link between quantum coherence and heat. The results provided by Eqs.~(\ref{equivalencework}) and (\ref{fluxes}) motivate a further analysis of the quantum coherence term to better understand its role in the first and second laws of thermodynamics in the quantum realm.

In the following section, we present a detailed description of perturbation theory in the interaction picture, which serves as the primary tool for obtaining up to second-order corrections to the thermodynamic quantities given by Eqs.~\eqref{work}, \eqref{heat}, and \eqref{coherence}. Such corrections allow us to 
decompose the coherence contribution given by Eq.~\eqref{coherence} into work and heat, proving that the first law is composed solely of such energetic quantities and no third term is needed. 

\section{Perturbative model}

In this section, we employ textbook perturbative theory \cite{Reis2025} to expand the density operator and the transition coefficients up to the second order. This framework allows us to write a perturbative quantum work, heat, and coherence, yielding to a perturbative version of the first law of quantum thermodynamics. In particular, our approach reveals that the quantum coherence energetic contribution is nothing more than work and heat energy exchanges. More details on the derivation of first- and second-order work are given in Appendices \ref{ap1} and \ref{apena}.

\subsection{Interaction picture}

The dynamics of quantum systems subjected to weak perturbative fields can be described through the decomposition of the total Hamiltonian
\begin{equation}
\hat H(t) = \hat H_0 + \hat V(t),
\end{equation}
where $\hat H_0$ denotes the free, time-independent part of the system, and $\hat V(t)$ accounts for the time-dependent external perturbation. The unperturbed Hamiltonian naturally defines the energy eigenbasis of the system, spanned by the eigenvectors $\{|n\rangle\}$, such that
\begin{equation}
\hat H_0 = \sum_n E_n |n\rangle\langle n|
\qquad \text{where}\qquad
\sum_n |n\rangle\langle n| = \mathbbm{1}.
\end{equation}
This basis plays a central role in the thermodynamic analysis, since variations in the eigenvalues $E_n(t)$ appear directly in the expression for the quantum work $W(t)$ defined by Eq.~\eqref{work}.

The initial state of the system is described by the density operator $\hat\rho(0)$, which is assumed to be diagonal in the basis $\{|k\rangle\}$
\begin{equation}\label{fkfjfhjf}
\hat\rho(0) = \sum_k \rho_k^{(0)} |k\rangle\langle k|.
\end{equation}
This basis is orthonormal and complete, such that $\sum_k |k\rangle\langle k| = \mathbbm{1}$; however, in general, it does not coincide with the energy basis $\{|n\rangle\}$. The difference between the two is responsible for the emergence of initial coherences when the state is rewritten in the energy basis \cite{PhysRevE.102.062152}. Indeed, the initial populations projected onto $|n\rangle$ are given by
\begin{equation}\label{vfkieik}
\rho_n^{(0)} = \langle n | \hat\rho(0) | n \rangle
= \sum_k \rho_k^{(0)} |\langle n|k\rangle|^2.
\end{equation}
This difference between the bases is crucial in the context of quantum thermodynamics, as it contributes to the coherence term $C(t)$, Eq.~\eqref{coherence}, that appears in the decomposition of the variation of the internal energy \cite{PhysRevE.102.062152}. {It is important to remember that in the interaction picture, neither basis are time-independent — all time evolution is encoded in the operator $\hat{U}(t)$.}

In order to describe the time evolution generated solely by the perturbation $\hat V(t)$, we adopt the interaction picture, where the dynamical contributions associated with $\hat H_0$ are removed via a unitary transformation \cite{Reis2025}. In this representation, the perturbation takes the form:
\begin{equation}
\hat V_I(t) = e^{i\hat H_0 t/\hbar}\, \hat V(t)\, e^{-i\hat H_0 t/\hbar}.
\end{equation}
In the interaction picture, the state $|\psi\rangle$ evolves according to the propagator $\hat U_I(t)$, defined as:
\begin{equation}\label{ofjjff}
|\psi_I(t)\rangle = \hat U_I(t)\,|\psi_I(0)\rangle.
\end{equation}
This operator satisfies the following differential equation:
\begin{equation}\label{dwede3221q2q}
i\hbar \frac{d\hat U_I(t)}{dt} = \hat V_I(t)\, \hat U_I(t),
\qquad
\hat U_I(0) = \mathbbm{1}.
\end{equation}
Integrating the equation above yields:
\begin{align}
\hat{U}_I(t) &= \mathbbm{1}-\left(\frac{i}{\hbar}\right)  \int_{0}^{t}\hat{V}_I(t')\hat{U}_I(t')\,dt',
\end{align}
which gives rise to the Dyson series \cite{Reis2025} for the propagator, expressed as:
\begin{align}
\nonumber\hat{U}_I(t) &=
\hat U_I^{(0)}(t)\;+\;\hat U_I^{(1)}(t)\;+\;\hat U_I^{(2)}(t) + \cdots\\
\nonumber &=\mathbbm{1}
  -\left(\frac{i}{\hbar}\right)\!\int_{0}^{t} dt'\,\hat{V}_I(t')+\\
  & \left(\frac{i}{\hbar}\right)^{\!2}\!\int_{0}^{t} dt' \,\hat{V}_I(t')\!\int_{0}^{t'} dt''\,
      \hat{V}_I(t'') -
      \cdots
\end{align}
For a comprehensive discussion of the interaction picture, the reader is referred to Ref. \cite{Reis2025}. Within this framework, the subsequent sections are devoted to the perturbative expansions of heat, work, and coherence, naturally leading to a ``perturbative version" of the first law in the context of quantum thermodynamics.

\subsection{Density Operator}

Starting from Eq.~\eqref{ofjjff}, the time evolution of the density operator in the interaction picture can be written as
\begin{align}\label{eewdwdw}
\hat{\rho}_I(t)=\hat{U}_I(t)\,\hat{\rho}_I(0)\,\hat{U}_I^{\dagger}(t).
\end{align}
Accordingly, the density operator can likewise be expressed in perturbative form as
\begin{align}\label{odpp}
\nonumber \hat{\rho}_{I}(t) &= \hat\rho^{(0)}+\hat\rho^{(1)}(t)+\hat\rho^{(2)}(t)+\cdots \\
\nonumber &=\hat{\rho}(0)
   - \left(\frac{i}{\hbar}\right)\int_{0}^{t} dt'\,
     \bigl[\hat{V}_{I}(t'),\,\hat{\rho}_{I}(0)\bigr] +\\
 &\left(\frac{i}{\hbar}\right)^{2}
\int_{0}^{t}\!dt' \int_{0}^{t'}\!dt'' \,
\bigl[\hat V_I(t'),\,[\hat V_I(t''),\hat\rho_I(0)]\bigr]-\cdots
\end{align}
Note that the initial density operator is denoted by $\hat{\rho}^{(0)}=\hat{\rho}(0)=\sum_k \rho_k^{(0)} |k\rangle\langle k|$, where $\{|k\rangle\}$ is the basis of the density operator, as defined in Eq.~\eqref{fkfjfhjf}. In the definitions of heat $Q(t)$, work $W(t)$, and coherence $C(t)$ given in Ref. \cite{PhysRevE.102.062152}, the diagonal matrix elements of the density operator $\hat\rho$, expressed in the basis $\{|k\rangle\}$, appear explicitly. For this reason, the relevant elements are
\begin{align} \label{rho_mat_elem}
\nonumber\rho_{k}(t) &=\langle k|\hat \rho_{I}(t)|
k\rangle \\
\nonumber&= \langle k| (\hat\rho_{I}^{(0)}+\hat\rho_{I}^{(1)}(t)+\hat\rho_{I}^{(2)}(t))|
k\rangle\\
&= \rho_{k}^{(0)}+\rho_{k}^{(1)}(t)+\rho_{k}^{(2)}(t)
\end{align}

\subsection{Transition coefficients}

In order to compute the heat $Q(t)$, work $W(t)$, and coherence $C(t)$ as defined in Ref. \cite{PhysRevE.102.062152}, the transition coefficients between the bases ${|n\rangle}$ and ${|k\rangle}$ are also required, as introduced in Eq.~\eqref{vfkieik}. These coefficients follow from Eqs.~\eqref{fkfjfhjf} and \eqref{eewdwdw}, where the density operator $\hat{\rho}_I(t)$ is written in the energy basis ${|n\rangle}$
\begin{align}
\langle n|\hat \rho_I(t)|n\rangle 
&= \sum_k \rho_k\,\langle n|\hat U_I(t)|k\rangle\,\langle n|\hat U_I(t)|k\rangle^*\\
&=\sum_k \rho_k\,|c_{nk}(t)|^2,
\end{align}
where
\begin{align}
c_{nk}(t) &= \langle n|\hat U_I(t)|k\rangle.
\end{align}
Since $c_{nk}(t)$ depends on the propagator $U_I(t)$, we can therefore expand it as
$c_{nk}(t) = c_{nk}^{(0)} + c_{nk}^{(1)}(t) + \cdots$, where
\begin{align}
c_{nk}^{(0)} = \langle n\,|\,k\rangle
\end{align}
\begin{align}\label{eerere}
c_{nk}^{(1)}(t) =-
\frac{i}{\hbar}\int_{0}^{t}
      \langle n|\hat V_I(t')|k\rangle\,dt'.
\end{align}
In the course of this analysis, the squared modulus of these coefficients will play a useful role
\begin{align} \label{c_n,k_exp}
\nonumber\big|c_{nk}(t)\big|^{2}
&=\big|c^{(0)}_{nk}+c^{(1)}_{nk}(t)\big|^{2}
\\&=\big|c^{(0)}_{nk}\big|^{2}
+2\,\mathrm{Re}\left\{\left[c^{(0)}_{nk}\right]^*c^{(1)}_{nk}(t)\right\}
+\big|c^{(1)}_{nk}(t)\big|^{2},
\end{align}
where the first term on the right-hand side is of zeroth order in the perturbation, the second term is of first order in the perturbation, and the third term is of second order in the perturbation.


\subsection{Perturbative quantum work}

We now turn to the quantum work, as developed in Ref. \cite{PhysRevE.102.062152} and presented in Eq.~\eqref{work}. In this equation, the quantities $\rho_{k}(t)$ and $|c_{nk}(t')|^{2}$ are defined in Eqs.~\eqref{rho_mat_elem} and \ref{c_n,k_exp}, respectively. Our goal is to expand the work in the form $W(t)=
W^{(0)}(t)\;+\;W^{(1)}(t)\;+\;W^{(2)}(t)\;+\;\cdots$, where $W^{(0)}(t)$ is of zeroth order, $W^{(1)}(t)$ is of first order, and $W^{(2)}(t)$ is of second order in the perturbation. The equation above can then be rewritten as
\begin{widetext}
\begin{equation}
W(t)=
\int_{0}^{t}\,dt' \sum_{n,k}
      \left[\rho_{k}^{(0)}+\rho_{k}^{(1)}(t')+\rho_{k}^{(2)}(t')\right]\,\left[\big|c^{(0)}_{nk}\big|^{2}
+2\,\mathrm{Re}\left\{\left[c^{(0)}_{nk}\right]^*c^{(1)}_{nk}(t')\right\}
+\big|c^{(1)}_{nk}(t')\big|^{2}\right]\,
      \frac{dE_{n}(t')}{dt'}.
\end{equation}
\end{widetext}
The zeroth-order correction can be written as:
\begin{align}
\nonumber W^{(0)}(t)&=
\int_{0}^{t}\,dt' \sum_{n,k}
      \rho_{k}^{(0)}\,\big|\langle n\,|\,k\rangle\big|^{2}\,
      \frac{dE_{n}(t')}{dt'}\\\nonumber&=\sum_{n}\left[E_{n}(t)-E_n(0)\right]\sum_{k} \,\big|\langle n\,|\,k\rangle\big|^{2}\,\rho_{k}^{(0)}\\&=\sum_{n}\left[E_{n}(t)-E_n(0)\right]\langle n\,|\hat \rho^{(0)}|\,n\rangle
\end{align}
The zeroth-order term accounts for the work arising exclusively from the explicit time dependence of the energy levels, with the populations remaining frozen at their initial values. This contribution does not depend explicitly on the perturbation $\hat V(t)$ and vanishes for static energy spectra.

The first-order contribution to the work can be expressed as:
\begin{align}
\nonumber W^{(1)}(t)=&\int_{0}^{t}\,dt'  \sum_{n,k}
      \rho_{k}^{(1)}(t')\,\big|\langle n\,|\,k\rangle\big|^{2}\,
      \frac{dE_{n}(t')}{dt'}+\\&2
\int_{0}^{t}\,dt' \sum_{n,k}
      \rho_{k}^{(0)}
      \,\mathrm{Re}\left\{\langle k\,|\,n\rangle c^{(1)}_{nk}(t')\right\}
      \frac{dE_{n}(t')}{dt'}
\end{align}
which simplifies to (see Appendix \ref{ap1})
\begin{align}\label{w112v}
W^{(1)}(t)=
      -\frac{2i}{\hbar}\,
\int_{0}^{t}dt'\;
      \int_{0}^{t'}d t''\;\sum_k
      \bigl\langle k\bigl|\,
            \bigl[\hat{V}_I(t''),\hat{\rho}^{(0)}\bigr]\dot {\hat H}(t')
      \bigr|k\bigr\rangle
\end{align}
The above term contributes only when
(i) the energy levels vary explicitly in time, and (ii) \([\hat V_I,\hat\rho(0)]\neq 0\). If either of these conditions holds, a reversible work contribution emerges, as it can be retrieved through a simple phase inversion of the external field. 

The second-order correction to the work is given by
\begin{widetext}
\begin{equation}
W^{(2)}(t)
=\int_{0}^{t}\,dt' \sum_{n,k}
\Big[
\rho_k^{(2)}(t')\,\big|c^{(0)}_{nk}\big|^{2}
+2\,\rho_k^{(1)}(t')\,\text{Re}\left\{ \left[c^{(0)}_{nk}\right]^{*}\,c^{(1)}_{nk}(t')\right\}
+\rho_k^{(0)}\,\big|c^{(1)}_{nk}(t')\big|^{2}
\Big]\,
\frac{dE_n(t')}{dt'}
\end{equation}
\end{widetext}
which can be recast as (see Appendix \ref{apena})
\begin{align}\label{rerwrfgggg}
\nonumber W^{(2)}(t)
&=\int_{0}^{t}\!dt'\;\sum_{k}\;\langle k|\dot {\hat H}(t')|k\rangle\;\sum_{n}\;\rho_k^{(0)}|c^{(1)}_{nk}(t')|^{2}\\&=\int_{0}^{t}\!dt'\;
\sum_{k}
\langle k|\dot{\hat H}(t')|k\rangle\,
P_k(t'),
\end{align}
where we have defined
\begin{equation}
P_k(t)\equiv
\sum_{n}\rho_k^{(0)}\,
\lvert c^{(1)}_{nk}(t)\rvert^2,
\end{equation}
that is, the probability for the system, initially prepared in the state $\lvert k\rangle$, to undergo transitions to other states at first order. This expression makes explicit that the second-order contribution to the work is entirely controlled by the explicit time dependence of the Hamiltonian, through the diagonal term $\langle k|\dot{\hat H}(t)|k\rangle$. In particular, $W^{(2)}(t)$ vanishes if the Hamiltonian is static ($\dot{\hat H}=0$), even in the presence of coupling between states. The role of $P_k(t)$ is to quantify the instability induced by the perturbation in the initially occupied state $\lvert k\rangle$. Thus, the second-order work does not depend on the occupation of specific final states, but rather on the probability of transitions to states different from the initial ones, weighted by the corresponding initial populations $\rho_k^{(0)}$.

In order to explicitly investigate this result, we consider a harmonic perturbation of the form $V_{nk}(t)=v_{nk}e^{-i\omega t}$ substituted into Eq.~\eqref{eerere}, following Ref.~\cite{Reis2025}. Here, $V_{nk}(t)$ is expressed in the Schrödinger picture as $V^{S}_{nk}(t)=v_{nk}e^{-i\omega t}$, and consequently, in the interaction picture one obtains
\[
V^I_{nk}(t)
=e^{i\omega_{nk}t}V^{S}_{nk}(t)
= v_{nk}e^{-i(\omega-\omega_{nk})t}.
\]
The probability of transition from the initially occupied state $\lvert k\rangle$ can thus be expressed as \cite{Reis2025}
\begin{align}
P_k(t)
&=\sum_n \rho_k^{(0)}\,\lvert c^{(1)}_{nk}(t)\rvert^2\\
&=\rho_k^{(0)}
\sum_n
\frac{\lvert v_{nk}\rvert^2}{\hbar^2}\,
\frac{\sin^2\!\bigl[(\omega-\omega_{nk})\,t/2\bigr]}
     {[(\omega-\omega_{nk})/2]^2}.
\label{eq:Pk_full}
\end{align}
It is worth noting that this expression explicitly involves the factor $\lvert v_{nk}\rvert^2$, reflecting the quadratic nature of the coupling in the transition probability, while the dependence on \(n\) accounts for all transition channels accessible from the initially occupied state \(\lvert k\rangle\). In the long-time limit, making use of the identity
\[
\frac{\sin^2(x t)}{x^2}\xrightarrow[t\to\infty]{}\pi t\,\delta(x),
\]
which holds in the distributional sense, we obtain
\begin{align}
P_k(t)
\xrightarrow[t\to\infty]{}
\frac{2\pi t}{\hbar^2}\,
\rho_k^{(0)}
\sum_n \lvert v_{nk}\rvert^2\,
\delta\!\bigl(\omega-\omega_{nk}\bigr),
\label{eq:Pk_assint}
\end{align}
which agrees with the standard expression of Fermi’s golden rule for the transition probability out of the state \(\lvert k\rangle\).

This result demonstrates that, at second order in the perturbation, the work is governed by transition probabilities to states other than the initial ones, being jointly controlled by the explicit time variation of the Hamiltonian and the resonance conditions enforced by the delta function in the case of harmonic driving. Because this contribution depends solely on $\lvert v_{nk}\rvert^2$, it is insensitive to phase reversals of the external perturbation and thus has an irreversible nature.

\subsection{Perturbative Quantum Heat}

We now turn to the quantum heat, as discussed in Ref. \cite{PhysRevE.102.062152} and presented in Eq.~\eqref{heat}. Our goal is to expand the heat in the form $Q(t)=
Q^{(0)}(t)\;+\;Q^{(1)}(t)\;+\;Q^{(2)}(t)\;+\;\cdots$, where $Q^{(0)}(t)$ is of zeroth order, $Q^{(1)}(t)$ is of first order, and $Q^{(2)}(t)$ is of second order in the perturbation. By substituting Eqs.~\eqref{rho_mat_elem} and \eqref{c_n,k_exp} into Eq.~\eqref{heat}, we can rewrite it as
\begin{widetext}
    \begin{equation}
Q(t)=
\int_{0}^{t}\,dt'  \sum_{n,k}
      E_{n}(t')\,
      \left[\big|c^{(0)}_{nk}\big|^{2}
+2\,\mathrm{Re}\left\{\left[c^{(0)}_{nk}\right]^*c^{(1)}_{nk}(t')\right\}
+\big|c^{(1)}_{nk}(t')\big|^{2}\right]
      \,
      \frac{d}{dt'}\left[\rho_{k}^{(0)}+\rho_{k}^{(1)}(t')+\rho_{k}^{(2)}(t')\right].
\end{equation}
\end{widetext}
The zeroth-order contribution vanishes, $Q^{(0)}(t)=0$, because $d\rho_{k}^{(0)}/dt=0$. We next consider the first-order correction to the heat, given by
\begin{align}\label{heatQ1}
\nonumber Q^{(1)}(t)&=
\int_{0}^{t}\,dt' \sum_{n,k}
      E_{n}(t')\,
      \big|c^{(0)}_{nk}\big|^{2}
      \,
      \frac{d}{dt'}\rho_{k}^{(1)}(t')\\&= \int_{0}^{t}\, dt' \sum_{k} 
\langle k|\hat H(t')|k\rangle \,\dot{\rho}^{(1)}_{k}(t')  .
\end{align}
Nevertheless, Eqs.~\eqref{rho11b} and \eqref{rho11} show that $\rho_k^{(1)}(t)=0$ for all $k$ and $t$, implying that $Q^{(1)}(t)=0$. The first nonzero contribution to the heat appears only at second order in the perturbation and is given by
\begin{align}\label{ptfmkff}
\nonumber Q^{(2)}(t)&=
\int_{0}^{t}\,dt' \sum_{n,k}
E_{n}(t')\,
\lvert c^{(0)}_{nk}\rvert^{2}
\,
\frac{d}{dt'}\rho_{k}^{(2)}(t')\\
&=
\int_{0}^{t}\,dt' \sum_{k}
\langle k|\hat H(t')|k\rangle \,
\dot{\rho}^{(2)}_{k}(t').
\end{align}

The population dynamics at second order can be written explicitly in terms of the first-order transition probabilities, according to Eq.~\eqref{eq2222}. Upon substituting this result into Eq.~\eqref{ptfmkff} and reorganizing the terms, we obtain
\begin{equation}\label{Q2-final}
Q^{(2)}(t)
=
\int_0^t dt'
\sum_{n,k}
\rho^{(0)}_k
\big[
E_k(t') - E_n(t')
\big]
\frac{d}{dt'}
\big|c^{(1)}_{nk}(t')\big|^2.
\end{equation}
This expression represents the central result for the second-order heat showing that this contribution is controlled by the energy differences between the levels linked by the transitions, as well as by the rate of increase of the transition probabilities, which also motivate us to relate the variation of the transition coefficients to the entropy flux as presented in Eq.~\eqref{heatcurrent}. In order to highlight the structural analogy with the second-order analysis of work, it is convenient to recast this expression in terms of transition rates. We thus define the elementary rate associated with the transition from the initially occupied state $\lvert k\rangle$ to a state $\lvert n\rangle$ as
\begin{equation}
R_{n\leftarrow k}(t)
\equiv
\frac{d}{dt}
\big|c^{(1)}_{nk}(t)\big|^2 .
\end{equation}
With this definition, Eq.~\eqref{Q2-final} can be equivalently written as
\begin{equation}
Q^{(2)}(t)
=
\int_0^t dt'
\sum_{n,k}
\rho^{(0)}_k
\big[
E_k(t') - E_n(t')
\big]
\, R_{n\leftarrow k}(t').
\end{equation}
In this formulation, the heat is expressed as a sumation over elementary contributions associated with all transition processes from the initially occupied states. Each $k \to n$ transition contributes with an energy amount given by the difference $E_k - E_n$, and weighted by both the transition rate and by the initial population of the departing state. Consequently, the second-order heat is directly linked to the perturbation-induced instability of the initially occupied states, in complete analogy with the role of transition probabilities in the quantum work case.

We now explicitly consider a harmonic perturbation of the form $V_{nk}(t) = v_{nk} e^{-i\omega t}$. In this case, the first-order transition coefficients can be evaluated analytically:
\begin{equation}\label{dwedwedwedee44iog}
\big|c^{(1)}_{nk}(t)\big|^2
=
\frac{|v_{nk}|^2}{\hbar^2}
\frac{
\sin^2\!\big[(\omega-\omega_{nk})t/2\big]
}{
\big[(\omega-\omega_{nk})/2\big]^2
}.
\end{equation}
In the long-time limit, the time derivative of these probabilities approaches the asymptotic form of Fermi’s golden rule
\begin{equation}\label{rwoewfjwejfkcd}
R_{n\leftarrow k}
=
\frac{2\pi}{\hbar^2}
|v_{nk}|^2
\delta(\omega-\omega_{nk}),
\end{equation}
indicating that only resonant transitions effectively contribute in the steady-state regime. This result shows that the second-order heat can be interpreted as a sum over elementary transition processes from the initially occupied states, where each transition $k \to n$ transfers a fixed amount of energy equal to $E_k - E_n$, occurring at a constant rate determined by Fermi’s golden rule. As in the case of work, this contribution is quadratic in the coupling, insensitive to phase reversal of the external perturbation, and thus exhibits an irreversible nature.


\subsection{Perturbative Quantum Coherence}

In this section, we explore quantum coherence, as discussed in Ref. \cite{PhysRevE.102.062152} and presented in Eq.~\eqref{coherence}. Our goal is to expand the coherence in the form $C(t)=
C^{(0)}(t)\;+\;C^{(1)}(t)\;+\;C^{(2)}(t)\;+\;\cdots$, where $C^{(0)}(t)$ is of zeroth order, $C^{(1)}(t)$ is of first order, and $C^{(2)}(t)$ is of second order in the perturbation. Equation \eqref{coherence} can then be rewritten as
\begin{widetext}
    \begin{align}
C(t) &=
\int_{0}^{t}\,dt' \sum_{n,k}
      E_{n}(t')\,\left[\rho_{k}^{(0)}+\rho_{k}^{(1)}(t')+\rho_{k}^{(2)}(t')\right]\,
      \frac{d}{dt'}\!\left[\big|c^{(0)}_{nk}\big|^{2}
+2\,\mathrm{Re}\left\{\left[c^{(0)}_{nk}\right]^*c^{(1)}_{nk}(t')\right\}
+\big|c^{(1)}_{nk}(t')\big|^{2}\right].
\end{align}
\end{widetext}

The zeroth-order correction is $C^{(0)}(t)=0$, since $d\lvert c^{(0)}_{nk}\rvert^{2}/dt=0$. The first-order correction can be written as
\begin{equation}
C^{(1)}(t)=\int_{0}^{t}\,dt' \sum_{n,k} E_n(t')\,\rho_k^{(0)}\,
\frac{d}{dt'}\Big[\,2\,\text{Re}\left\{[c^{(0)}_{nk}]^{*}c^{(1)}_{nk}(t')\right\}\Big].
\label{eq:c1}
\end{equation}
By integrating the equation above by parts, we obtain
\begin{equation}\label{wwewewewe}
C^{(1)}(t) = q^{(1)}(t)+w^{(1)}(t)
\end{equation}
where
\begin{align}
\nonumber q^{(1)}(t)&=\sum_{n,k}\rho_k^{(0)}\,E_n(t)\;
2\text{Re}\left\{[c^{(0)}_{nk}]^{*}c^{(1)}_{nk}(t)\right\}\\&=-\,\frac{i}{\hbar}\,\int_{0}^{t} dt' \sum_{k}\;
\langle k|\,[\,\hat V_I(t'),\,\hat\rho^{(0)}\,]\,H(t)\,|k\rangle ,
\end{align}
constitutes the first-order contribution to the coherent heat, and
\begin{align}
\nonumber w^{(1)}(t)&=-\int_{0}^{t}\,dt' \sum_{n,k}\rho_k^{(0)}\,\dot E_n(t')\;
2\text{Re}\!\left\{[c^{(0)}_{nk}]^{*}c^{(1)}_{nk}(t')\right\} \\&=\frac{i}{\hbar}\,\int_{0}^{t} dt'\,\int_{0}^{t'} dt''\;
\sum_{k} \langle k|\,\big[\,\hat V_I(t''),\,\hat\rho^{(0)}\,\big]\;\dot {\hat H}(t')\,|k\rangle
\end{align}
represents the first-order “coherent work”. Such decomposition reinforces the work definition in Refs. \cite{alipour2022entropy,ahmadi2023contribution} due to the eigenvectors contribution, while the transition coefficients variation also contributes as quantum heat, allowing it to compose an entropy flux analogously to Eq.~\eqref{heatcurrent}. In the above, we used Eqs.~\eqref{plmju}, \eqref{rrefff}, and \eqref{rrefff2}. The term $C^{(1)}(t)$ describes the initial emergence of coherences in the system, that is, the generation of quantum superpositions between distinct states induced by the action of the external perturbation. While the populations remain unchanged at first order, coherences can be created immediately through the interaction, since the coupling with the external field breaks the purely diagonal structure of the initial state. This mechanism reflects the ability of the field to induce quantum correlations and relative phases already in the linear regime of the perturbation, even in the absence of population redistributions.

The decomposition of coherence into coherent heat and coherent work (Eq. \eqref{wwewewewe}) allows for a more refined physical interpretation of this process. The term $q^{(1)}(t)$ is associated with the coherent contribution to the heat, corresponding to the energy change arising from the modification of the system state in the presence of a fixed Hamiltonian. Its structure depends explicitly on the commutator between the interaction and the initial state, indicating that this energy exchange occurs exclusively due to the generation of quantum coherences, without being related to explicit changes in the system’s energy spectrum. Thus, $q^{(1)}(t)$ represents a genuinely quantum form of heat, absent in classical or fully incoherent regimes. In turn, the term $w^{(1)}(t)$ can be interpreted as the first-order coherent work, having a structure similar to the quantum work (Eq.~\eqref{w112v})—up to a prefactor.

The second-order contribution correction is written as
\begin{equation}\label{45fef}
C^{(2)}(t)
=\int_0^t\,dt' \sum_{n,k}
E_n(t')\,\rho_k^{(0)}\,
\frac{d}{dt'}\big|c^{(1)}_{nk}(t')\big|^2 ,
\end{equation}
where the term $\lvert c^{(1)}_{nk}(t)\rvert^2$ denotes the transition probability arising from the first-order generated coherences. Upon integrating Eq.~\eqref{45fef} by parts and applying the initial condition $c^{(1)}_{nk}(0)=0$, we obtain
\begin{equation}
C^{(2)}(t)=q^{(2)}(t)+w^{(2)}(t),
\label{eq:C2-decomp}
\end{equation}
where
\begin{equation}
q^{(2)}(t)
=
\sum_{n,k}\rho_k^{(0)}\,E_n(t)\,
\big|c^{(1)}_{nk}(t)\big|^2 ,
\label{eq:q2}
\end{equation}
is the coherent heat associated with the quantum coherence variation, which emerges from the integration over the interval $[0,t]$ where we used the initial condition given above, and
\begin{equation}
w^{(2)}(t)
=
-\sum_{n,k}\rho_k^{(0)}
\int_0^t\,dt'
\frac{dE_n(t')}{dt'}\,
\big|c^{(1)}_{nk}(t')\big|^2 .
\label{eq:w2}
\end{equation}
Equation \eqref{eq:q2} describes the energy contribution associated with coherent transition probabilities, weighted by the corresponding energy levels. Since there is no explicit dependence on the time derivative of the Hamiltonian, this term is associated with the effective redistribution of populations induced by quantum coherence and can be interpreted as second-order coherent heat. In contrast, Eq. \eqref{eq:w2} depends explicitly on the temporal variation of the Hamiltonian eigenvalues, characterizing the energy transferred due to the perturbation in the presence of coherent transitions, and thus corresponds to second-order coherent work. Equations~\eqref{eq:c1}--\eqref{eq:w2} highlight that, while first-order coherence reflects only the initial generation of quantum superpositions, second-order coherence quantifies how these superpositions begin to control effective transition probabilities, giving rise to coherent energy contributions analogous to heat and work.

\subsection{Perturbative Quantum first law}

In the quantum regime, the operational form of the first law of thermodynamics can be expressed in the extended form as
\begin{equation}
U(t)=Q(t)+W(t)+C(t),
\end{equation}
where, in addition to the usual contributions of heat \(Q(t)\) and work \(W(t)\), the term
\(C(t)\) represents the energy contribution associated with the generation of quantum coherences in the system’s state. This extended form of the first law reflects the fact that, in quantum systems subject to an external perturbation, coherent dynamics can contribute independently to the variation of internal energy. When the dynamics is treated perturbatively, all the thermodynamic quantities admit an order-by-order expansion in the interaction picture, which implies a corresponding decomposition of the internal energy,
\begin{equation}
U(t)= U^{(0)}(t)+ U^{(1)}(t)+ U^{(2}(t) + \cdots,
\end{equation}
with
\begin{equation}
U^{(n)}(t)=Q^{(n)}(t)+W^{(n)}(t)+C^{(n)}(t),
\end{equation}
at each order of the perturbation. This formulation highlights that quantum coherence acts as an additional channel of energy exchange, complementary to the classical mechanisms of heat and work, and becomes essential for a consistent description of the energy balance in regimes where coherent dynamics play an active role.

\subsubsection{Zeroth order}
To zeroth order in the perturbation, the system’s dynamics produce neither transitions nor quantum coherences. The zeroth-order work is directly given by
\begin{equation}
W^{(0)}(t)
=
\sum_n \big[ E_n(t) - E_n(0) \big]\,
\langle n | \hat{\rho}^{(0)} | n \rangle .
\end{equation}
This indicates that the variation of the internal energy is solely due to the explicit time dependence of the energy levels, weighted by the initial populations. As the state remains diagonal in the energy basis and the occupations remain unchanged at this order, no population redistribution or coherence generation occurs. Therefore, both the heat and the coherent contribution vanish,
\begin{equation}\label{feewfweffwwedew}
Q^{(0)}(t)=0,
\qquad
C^{(0)}(t)=0.
\end{equation}

Accordingly, the extended first law at zeroth order can be written as
\begin{equation}
U^{(0)}(t)=W^{(0)}(t),
\end{equation}
indicating that the entire change in internal energy is solely attributed to the work performed by the external control on the Hamiltonian. Physically, the zeroth order describes a purely adiabatic regime, in which the quantum state is rigidly transported by the evolution of the energy spectrum, without any dissipative or coherent effects.

\subsubsection{First order}

To first order in the perturbation, the variation of internal energy is influenced by the generation of quantum coherences, while the energy level populations remain unchanged. The first-order work is provided by Eq.~\eqref{w112v},
\begin{equation}\label{ororkrmr}
W^{(1)}(t)
=
-\frac{2i}{\hbar}
\int_0^t dt'
\int_0^{t'} dt''
\sum_k
\langle k|
[\hat V_I(t''),\hat\rho^{(0)}]\,
\dot{\hat H}(t')
|k\rangle ,
\end{equation}
indicating that this contribution arises explicitly from the noncommutativity between the initial state and the interaction, as well as from the time dependence of the Hamiltonian. Consequently, it represents a coherent contribution. The first-order heat is given by Eq.~\eqref{heatQ1}. Nonetheless, because the first-order population correction vanishes,
$\rho_k^{(1)}(t)=0$,
no population redistribution occurs between the energy levels, which directly leads to:
\begin{equation}
Q^{(1)}(t)=0.
\end{equation}
At this order, coherence emerges explicitly and can be derived by integrating by parts the terms corresponding to the amplitude dynamics. The resulting expression can be written as
\begin{equation}
C^{(1)}(t)=q^{(1)}(t)+w^{(1)}(t),
\end{equation}
where the term
\begin{equation}\label{dwdewdwefrfgg}
q^{(1)}(t)
=
-\frac{i}{\hbar}
\int_0^t dt'\,
\sum_k
\langle k|
[\hat V_I(t'),\hat\rho^{(0)}]\,
\hat H(t)
|k\rangle
\end{equation}
represents the \emph{first-order coherent heat}, since it is associated exclusively with the modification of the quantum state, without involving the explicit variation of the Hamiltonian. The term
\begin{equation}
w^{(1)}(t)
=
\frac{i}{\hbar}
\int_0^t dt'
\int_0^{t'} dt''
\sum_k
\langle k|
[\hat V_I(t''),\hat\rho^{(0)}]\,
\dot{\hat H}(t')
|k\rangle
\end{equation}
represents the \emph{first-order coherent work}, structurally similar to the quantum work \(W^{(1)}(t)\). Accordingly, the first law at first order can be written as
\begin{equation}\label{dwwdwwwwww}
U^{(1)}(t)
=
W^{(1)}(t)
+
C^{(1)}(t)=\left[W^{(1)}(t)
+
w^{(1)}(t)\right]+q^{(1)}(t)
\end{equation}

From a physical standpoint, the first-order correction demonstrates that, even without population-mediated heat exchanges, the system’s internal energy is affected by the generation of quantum coherences. These coherences give rise to inherently quantum contributions to the energy balance, appearing as coherent components of heat and work in the linear perturbation regime.

\subsubsection{Second order}

At second order in the perturbation, the change in internal energy becomes governed by the transition probabilities induced by the first-order generated quantum coherences. Unlike the linear regime, where coherences appear only as superpositions, at second order they effectively control the system’s energy redistribution, giving rise to contributions analogous to heat and work. The second-order coherent contribution is obtained by integrating Eq.~\eqref{45fef} by parts and using the initial condition $c^{(1)}_{nk}(0)=0$, resulting in the decomposition
\begin{equation}
C^{(2)}(t)=q^{(2)}(t)+w^{(2)}(t).
\end{equation}
The term
\begin{equation}\label{calorcoerente}
q^{(2)}(t)
=
\sum_{n,k}\rho_k^{(0)}\,E_n(t)\,
\big|c^{(1)}_{nk}(t)\big|^2
\end{equation}
denotes the energy contribution arising from the coherent transition probabilities, weighted by the corresponding occupied energy levels. As it does not explicitly depend on the time derivative of the Hamiltonian, this term is linked to the effective population redistribution induced by quantum coherence and may be interpreted as \emph{second-order coherent heat}. In contrast, we obtain the term
\begin{align}\label{trabalhocoerente}
\nonumber w^{(2)}(t)
&=
-\sum_{n,k}\rho_k^{(0)}
\int_0^t\,dt'
\frac{dE_n(t')}{dt'}\,
\big|c_{nk}^{(1)}(t')\big|^2
\\&=
-
\int_0^t dt'
\sum_n
\langle n|\dot{\hat H}(t')|n\rangle\,
\tilde P_n(t')
\end{align}
where
\begin{align}
\tilde P_n(t)
=
\sum_k \rho_k^{(0)}\,\big|c_{nk}^{(1)}(t)\big|^2
\end{align}
quantifies the coherent probability of occupying the state $|n\rangle$ starting from the initially occupied states. This formulation makes explicit that $w^{(2)}(t)$ corresponds to the \emph{second-order coherent work}, describing the energy transferred by the external control to the coherently accessed states. Unlike the term $W^{(2)}(t)$, which is associated with the instability of the initially occupied states, $w^{(2)}(t)$ quantifies how this work is redistributed among the final states, highlighting that both represent complementary contributions of coherent dynamics to the energy balance. The second-order population contribution to the heat can be written as
\begin{equation}\label{dwedeeewwwweew}
Q^{(2)}(t)
=
\int_0^t dt'
\sum_{n,k}\rho_k^{(0)}
\big[E_k(t')-E_n(t')\big]\,
\frac{d}{dt'}\big|c^{(1)}_{nk}(t')\big|^2 .
\end{equation}
This expression shows that the second-order heat is governed by the energy differences between the states connected by the transitions and by the rate of change of the transition probabilities. Finally, the second-order work can be expressed as:
\begin{equation}\label{dedededededeed}
W^{(2)}(t)
=
\int_0^t dt'
\sum_k
\langle k|\dot{\hat H}(t')|k\rangle\,P_k(t'),
\end{equation}
where
\begin{equation}
P_k(t)
=
\sum_n\rho_k^{(0)}
\big|c^{(1)}_{nk}(t)\big|^2 .
\end{equation}
This formulation shows that the second-order work contribution is fully determined by the time dependence of the Hamiltonian, weighted by the perturbation-induced instability of the initially occupied states. Notably, $W^{(2)}(t)$ is zero for static Hamiltonians, even when couplings between states are present.

Combining these results, the second-order contribution to the first law can be written as
\begin{align}\label{rrrrrrrrr}
\nonumber U^{(2)}(t)
&=
W^{(2)}(t)
+
Q^{(2)}(t)
+
C^{(2)}(t)\\&=\left[W^{(2)}(t)
+
w^{(2)}(t)\right]+\left[Q^{(2)}(t)+q^{(2)}(t)\right]
\end{align}
highlighting that, at second order, quantum coherence acts as an active mechanism controlling the redistribution and modulation of the system’s energy, giving rise to well-defined coherent contributions to both heat and work.

\section{Application: a two-level system subjected to a harmonic perturbation}

We consider a two-level system, whose unperturbed Hamiltonian is diagonal in the energy basis,
\begin{equation}\label{fewfewweewe}
\hat H_0 =
\begin{pmatrix}
E_1 & 0 \\
0 & E_2
\end{pmatrix},
\qquad E_2 > E_1,
\end{equation}
with the corresponding eigenvectors denoted as
\begin{equation}
|1^{(0)}\rangle =
\begin{pmatrix}
1 \\ 0
\end{pmatrix},
\qquad
|2^{(0)}\rangle =
\begin{pmatrix}
0 \\ 1
\end{pmatrix}.
\end{equation}
This system serves as the minimal model for transitions induced by external fields and arises in several physical contexts, including two-level atoms, spins in magnetic fields, and qubits. At $t=0$, the system is exposed to a harmonic external perturbation, described by the operator
\begin{equation}
\hat V(t)
=
\varepsilon
\begin{pmatrix}
0 & e^{i\omega t} \\
e^{-i\omega t} & 0
\end{pmatrix},
\end{equation}
where $\varepsilon$ is the coupling strength and $\omega$ is the frequency of the perturbation. The total Hamiltonian of the system then becomes
\begin{equation}
\hat H(t)=\hat H_0+\hat V(t),
\end{equation}
thereby introducing an explicit coupling between the two energy levels.

We consider the two-level system governed by the unperturbed Hamiltonian in Eq.~\eqref{fewfewweewe}, initially prepared in a statistical state diagonal in the energy eigenbasis
\begin{equation}
\hat\rho^{(0)}
=
\rho_1^{(0)}|1^{(0)}\rangle\langle1^{(0)}|
+
\rho_2^{(0)}|2^{(0)}\rangle\langle2^{(0)}|,
\qquad
\rho_1^{(0)}+\rho_2^{(0)}=1.
\end{equation}
Here, no initial coherences are present. The ensuing dynamics are fully determined by the harmonic perturbation, which can drive transitions between the energy levels and create quantum superpositions as the system evolves. In the subsequent sections, this model will provide an explicit example to demonstrate the application of the general expressions derived in this work, enabling a concrete analysis of harmonic-field-induced dynamics in a fundamental quantum system.

\subsection{Zeroth order}

At zeroth order in the perturbation, the dynamics of the two-level system do not induce transitions nor generate quantum coherences. The system’s state remains diagonal in the eigenbasis of $\hat H_0$, with constant populations
$\rho_n^{(0)}$. Therefore, the zeroth-order work is $W^{(0)}(t)=0$, since
$E_n(t)=E_n(0)$. Similarly, according to Eq.~\eqref{feewfweffwwedew}, we have $Q^{(0)}(t)=0$ and $C^{(0)}(t)=0$. The internal energy change at zeroth order is thus:
$U^{(0)}(t)=0$, indicating that, in this regime, there is no energy exchange with the environment nor internal energy redistribution. Physically, the zeroth order describes a purely adiabatic regime, in which the system is rigidly carried along by the temporal evolution of the energy spectrum, without changes in the populations or the generation of quantum superpositions. In this limit, the influence of the external field has not yet manifested in the dynamics, and the system’s entire energy structure remains unaltered. 

\subsection{First order}

The system is driven by a purely off-diagonal harmonic perturbation,
\begin{equation}
\hat V(t)
=
\varepsilon
\begin{pmatrix}
0 & e^{i\omega t}\\
e^{-i\omega t} & 0
\end{pmatrix}.
\end{equation}
Within the interaction picture defined by $\hat H_0$, the perturbation is expressed as
\begin{equation}
\hat V_I(t)
=
\varepsilon
\begin{pmatrix}
0 & e^{i(\omega-\omega_{21})t}\\
e^{-i(\omega-\omega_{21})t} & 0
\end{pmatrix},
\end{equation}
where $\hbar\,\omega_{21}=E_2-E_1>0$. As the initial state is diagonal, the commutator of the interaction with the density operator can be written explicitly as
\begin{equation}
[\hat V_I(t),\hat\rho^{(0)}]
=
\varepsilon(\rho_2^{(0)}-\rho_1^{(0)})
\begin{pmatrix}
0 & e^{i(\omega-\omega_{21})t}\\
- e^{-i(\omega-\omega_{21})t} & 0
\end{pmatrix},
\label{commutator_example}
\end{equation}
and is therefore purely off-diagonal. The term $q^{(1)}(t)$ which contributes to the first-order coherent contribution, is given by Equation \ref{dwdewdwefrfgg}. Since $\hat H_0$ is diagonal and the commutator~\eqref{commutator_example} has no diagonal elements, it immediately follows that $q^{(1)}(t)=0$. The first-order quantum work is given by Eq.~\eqref{ororkrmr}. For this calculation, the time derivative of the total Hamiltonian is required.
\begin{equation}
\dot{\hat H}(t')
=
\dot{\hat V}(t')
=
i\varepsilon\omega
\begin{pmatrix}
0 & e^{i\omega t'} \\
- e^{-i\omega t'} & 0
\end{pmatrix}.
\label{eq:Hdot_exemplo}
\end{equation}
The matrix product in the definition of $W^{(1)}(t)$ is, therefore
\begin{align}
\nonumber &[\hat V_I(t''),\hat\rho^{(0)}]\,
\dot{\hat H}(t')
=\\&
i\varepsilon^2\omega\big(\rho_1^{(0)}-\rho_2^{(0)}\big)
\begin{pmatrix}
e^{i[(\omega-\omega_{21})t''-\omega t']} & 0 \\
0 & e^{i[-(\omega-\omega_{21})t''+\omega t']} 
\end{pmatrix}.
\label{eq:produto_matricial}
\end{align}
Since the integrand of $W^{(1)}(t)$ involves the trace of this operator, we need to sum the diagonal elements of Eq.~\eqref{eq:produto_matricial}, yielding
\begin{align}
\nonumber &\mathrm{Tr}\!\left\{[\hat V_I(t''),\hat\rho^{(0)}]\,
\dot{\hat H}(t')\right\}
\\&=
i\varepsilon^2\omega\big(\rho_1^{(0)}-\rho_2^{(0)}\big)
\Big[
e^{i[(\omega-\omega_{21})t''-\omega t']}
+
e^{i[-(\omega-\omega_{21})t''+\omega t']}
\Big].
\label{eq:integrando_bruto}
\end{align}
This trace is precisely the integrand in the time integrals defining the first-order quantum work, so that, upon substitution of Eq.~\eqref{eq:integrando_bruto} in \eqref{dwdewdwefrfgg}, we obtain:
\begin{widetext}
\begin{equation}
W^{(1)}(t)
=
-\frac{2i}{\hbar}
\int_0^t dt'
\int_0^{t'} dt''\,
i\varepsilon^2\omega\big(\rho_1^{(0)}-\rho_2^{(0)}\big)
\Big[
e^{i[(\omega-\omega_{21})t''-\omega t']}
+
e^{i[-(\omega-\omega_{21})t''+\omega t']}
\Big],
\end{equation}
\end{widetext}
which will now be integrated step by step over the variables $t''$ e $t'$. After performing the integration, we obtain
{\small \begin{equation}\label{fgrkoo}
W^{(1)}(t)
=
\frac{4\,\varepsilon^{2}\,\omega\,
\big(\rho_1^{(0)}-\rho_2^{(0)}\big)}
{\hbar\,(\omega-\omega_{21})}
\left[
\frac{1-\cos(\omega t)}{\omega}
-
\frac{1-\cos(\omega_{21} t)}{\omega_{21}}
\right].
\end{equation}}From the general definitions of the perturbative formalism, it is observed that the terms
$W^{(1)}(t)$ and $w^{(1)}(t)$ have exactly the same integrand, differing only by the prefactor accompanying each contribution. While the first-order quantum work arises with the factor $-2i/\hbar$, the corresponding coherent term comes with the factor $+i/\hbar$. As a direct consequence of this structural difference, the general relation is obtained
\begin{equation}
w^{(1)}(t)=-\frac{1}{2}\,W^{(1)}(t).
\end{equation}
The first law of Thermodynamics, explicitly including the coherent contribution, is written as in Eq.~\eqref{dwwdwwwwww}
\begin{equation}
U^{(1)}(t)
=\left[W^{(1)}(t)
+
w^{(1)}(t)\right]+q^{(1)}(t).
\end{equation}
We know that $q^{(1)}(t)=0$, which leads to
\begin{equation}
U^{(1)}(t)
=\frac{1}{2}\,W^{(1)}(t)=-\,w^{(1)}(t).
\end{equation}
From a physical point of view, this result indicates that, at first order, the energy supplied by the external field is not converted into coherent heat. On the other hand, half of the first-order quantum work is temporarily stored as quantum coherence, as reflected in the term $w^{(1)}(t)$.

\subsection{Second order}

The second-order contribution for the quantum work is given by Eq.~\eqref{dedededededeed}, therefore, we need the matrix element $\langle k|\dot{\hat H}(t')|k\rangle$. In the given example, the total Hamiltonian is $\hat H(t)=\hat H_0+\hat V(t)$, with static $\hat H_0$  and, therefore, $\dot{\hat H}(t)=\dot{\hat V}(t)$, as presented in detail in Eq.~\eqref{eq:Hdot_exemplo}. Then, we have $\langle k|\dot{\hat H}(t')|k\rangle=0$ and, consequently, $W^{(2)}(t)=0$.

The quantum heat is given by Eq.~\eqref{dwedeeewwwweew}. In the given example, $E_1$ e $E_2$ are constants (static non-perturbed Hamiltonian),
then $E_k(t')-E_n(t')=E_k-E_n$. In this way, we have
\begin{equation}
Q^{(2)}(t)
=
\sum_{n,k}\rho_k^{(0)}(E_k-E_n)
\left[
|c_{nk}^{(1)}(t)|^{2}-|c_{nk}^{(1)}(0)|^{2}
\right].
\end{equation}
Since $c_{nk}^{(1)}(0)=0$, we can rewrite the above equation in the following way
\begin{equation}
Q^{(2)}(t)
=
\sum_{n,k}\rho_k^{(0)}(E_k-E_n)\,|c_{nk}^{(1)}(t)|^{2}.
\label{eq:Q2_border}
\end{equation}
For a two-level system, it implies that
\begin{equation}
Q^{(2)}(t)
=
\rho_1^{(0)}(E_1-E_2)\,|c_{21}^{(1)}(t)|^{2}
+
\rho_2^{(0)}(E_2-E_1)\,|c_{12}^{(1)}(t)|^{2}.
\label{eq:Q2_2lvl}
\end{equation}
Defining $\hbar\omega_{21}=E_2-E_1$, and using the result obtained from Eq.~\eqref{dwedwedwedee44iog} for an harmonic perturbation, we arrive at
\begin{equation}
|c_{21}^{(1)}(t)|^{2}=|c_{12}^{(1)}(t)|^{2}
=
\frac{\varepsilon^{2}}{\hbar^{2}}\,
\frac{\sin^{2}\!\left[(\omega-\omega_{21})t/2\right]}{[(\omega-\omega_{21})/2]^{2}}.
\label{eq:c1_sq_noDelta}
\end{equation}
Upon substitution of Eq.~\eqref{eq:c1_sq_noDelta} at \eqref{eq:Q2_2lvl}, we arrive at the following result
\begin{equation}
Q^{(2)}(t)
=\hbar\omega_{21}
\frac{\varepsilon^{2}}{\hbar^2}\,
\big(\rho_2^{(0)}-\rho_1^{(0)}\big)\,
\frac{\sin^{2}\!\left[(\omega-\omega_{21})t/2\right]}{[(\omega-\omega_{21})/2]^{2}}.
\label{eq:Q2_final_noDelta}
\end{equation}
Considering the asymptotic form of Fermi’s Golden Rule, Eq.~\eqref{rwoewfjwejfkcd}, we obtain for the second-order heat rate of change
\begin{equation}
\frac{Q^{(2)}(t)}{t}=\hbar\omega_{21}
\frac{\varepsilon^{2}}{\hbar^2}\,
\big(\rho_2^{(0)}-\rho_1^{(0)}\big)\,2\pi
\delta(\omega-\omega_{nk}),
\end{equation}
showing that, in the steady-state regime, only resonant transitions contribute to the second-order heat. 

The second-order perturbative coherence is written as $C^{(2)}(t)=q^{(2)}(t)+w^{(2)}(t)$, where the coherent heat is given in Eq.~\eqref{calorcoerente} and the coherent work in Eq.~\eqref{trabalhocoerente}. For the two-level system under the considered harmonic perturbation, we have $\langle n|\dot{\hat H}(t')|n\rangle=0$, considering Eq.~\eqref{eq:Hdot_exemplo}. Therefore, the second-order coherent work vanishes, $w^{(2)}(t)=0$. For the coherent heat, the expression can be simplified as
\begin{equation}
q^{(2)}(t)=\sum_{n,k}\rho_k^{(0)}\,E_n\,\big|c_{nk}^{(1)}(t)\big|^2,
\end{equation}
since $E_n(t)=E_n$. For a two-level system, we have
\begin{equation}
q^{(2)}(t)=
\rho_1^{(0)}\,E_2\,\big|c_{21}^{(1)}(t)\big|^2
+
\rho_2^{(0)}\,E_1\,\big|c_{12}^{(1)}(t)\big|^2.
\end{equation}
Using the result from Eq.~\eqref{eq:c1_sq_noDelta}, we obtain
\begin{equation}
q^{(2)}(t)=
\Big(\rho_1^{(0)}E_2+\rho_2^{(0)}E_1\Big)\,
\frac{\varepsilon^{2}}{\hbar^{2}}\,
\frac{\sin^{2}\!\left[(\omega-\omega_{21})t/2\right]}{[(\omega-\omega_{21})/2]^{2}}.
\end{equation}
In the asymptotic limit of Fermi’s Golden Rule, we then obtain the second-order coherent heat rate of change
\begin{equation}
\frac{q^{(2)}(t)}{t}=
\Big(\rho_1^{(0)}E_2+\rho_2^{(0)}E_1\Big)\,
\frac{\varepsilon^{2}}{\hbar^{2}}\,
2\pi
\delta(\omega-\omega_{nk}).
\end{equation}

The second-order correction to the first law was written in Eq.~\eqref{rrrrrrrrr} as
\begin{equation}
U^{(2)}(t)
=
\left[W^{(2)}(t)
+
w^{(2)}(t)\right]+\left[Q^{(2)}(t)+q^{(2)}(t)\right].
\end{equation}
For this example, the second-order work contribution is zero, and therefore we have
\begin{equation}
U^{(2)}(t)
=
Q^{(2)}(t)+q^{(2)}(t),
\end{equation}
which can be rewritten as
\begin{equation}
U^{(2)}(t)=
\Big(\rho_1^{(0)}E_1+\rho_2^{(0)}E_2\Big)\,
\frac{\varepsilon^{2}}{\hbar^{2}}\,
\frac{\sin^{2}\!\left[(\omega-\omega_{21})t/2\right]}{[(\omega-\omega_{21})/2]^{2}}.
\end{equation}

\subsection{Conclusion of the example}

In zeroth order, the Hamiltonian is static and the system’s state remains diagonal in the energy basis, with no transitions or generation of coherences. As a result, there is no energy exchange, and heat, work, and coherence contributions are all zero. At first order, the off-diagonal perturbation induces coherences while the level populations remain unchanged. In this regime, no population-mediated heat arises, and the energetic response is dominated by work, linked to the time dependence of the Hamiltonian and the energy temporarily stored in the form of coherence. This process is fundamentally reversible.

At second order, the previously generated coherences begin to determine the effective transition probabilities between levels. Because the Hamiltonian spectrum is time-independent, the work contributions vanish, and the internal energy change stems entirely from heat, which is quadratic in the coupling and characterizes an irreversible process governed by the transition rates. In the long-time limit, this heat is described by Fermi’s Golden Rule.

\begin{widetext}
\begin{figure*}[htb]
\centering
    \includegraphics[width=1.0\textwidth]{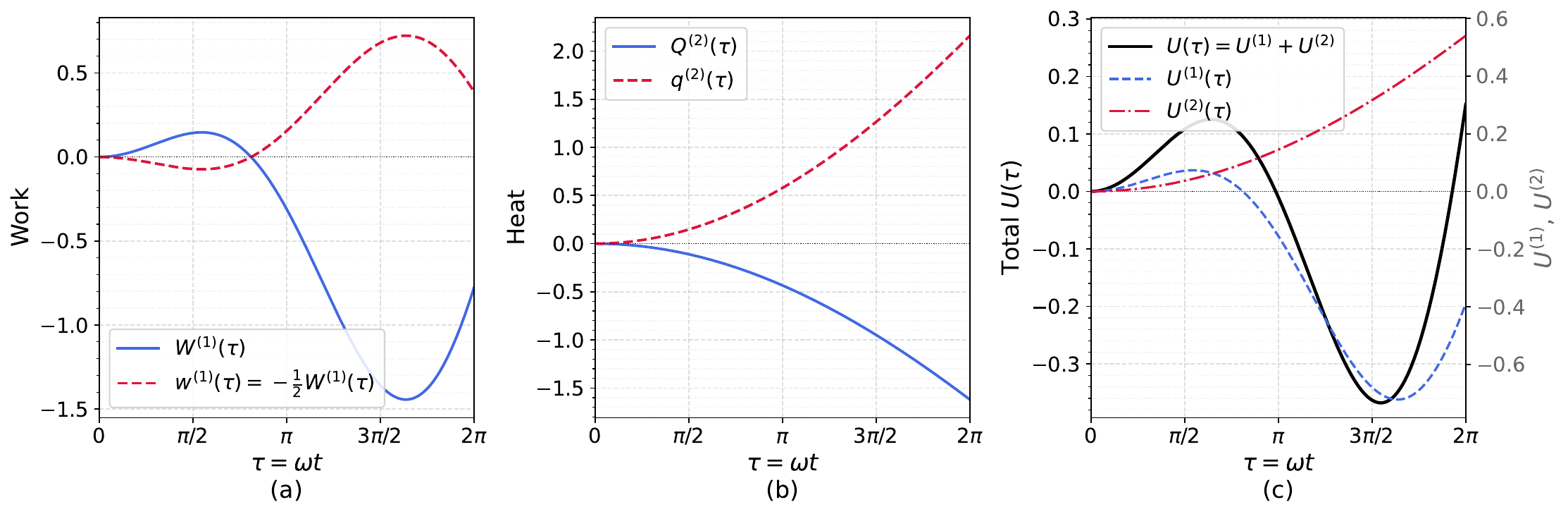}
    
    \caption{\justifying First- and second-order work, heat, and internal energy of our example. Parameters: $\hbar = 1$, $\varepsilon = 0.3$, $\rho_1^{(0)} = 0.8$, $\rho_2^{(0)} = 0.2$, $E_1 = 0$, $E_2 = \omega_{21}/\omega$, and $\omega_{21} = 1$.}
    \label{examplefig}
\end{figure*}
\end{widetext}

As can be seem in Fig. \ref{examplefig}, the temporal evolution of the energetic contributions in the presence of a harmonic perturbation reveals the distinct phenomenological roles played by quantum coherence and population dynamics. On short interaction timescales, the internal energy exchange presented on Fig. \ref{examplefig} \textcolor{blue}{(c)} is entirely governed by the first-order coherent work, $W^{(1)}(t)$ (Fig. \ref{examplefig} \textcolor{blue}{(a)}), this contribution displays fully reversible, oscillatory dynamics, characteristic of quantum superposition states, and proceeds without concurrent changes in the energy level populations. As the evolution proceeds, second-order contributions to the internal energy become dominant, and we observe the emergence of irreversible coherent heat currents, $Q^{(2)}(t)$ and $q^{(2)}(t)$ (Fig. \ref{examplefig} \textcolor{blue}{(b)}), driven by transition probabilities in agreement with Fermi’s golden rule. This representation unambiguously elucidates the crossover from a coherence-dominated quantum regime to a dissipative thermodynamic regime, indicating that quantum coherence functions as an initial transient energetic resource, after which irreversible thermalization predominates in determining the dynamics of the system.

\section{Concluding Remarks}
In this work, we investigated the first law of quantum thermodynamics within a time-dependent perturbative framework. Motivated by recent discussions concerning the thermodynamic role of quantum coherence, we derived perturbative corrections to work, heat, and coherence contributions up to second order, allowing a direct analysis of how coherent quantum dynamics affect energy exchanges in quantum systems.

Our results show that the energetic contribution associated with quantum coherence can be decomposed into coherent heat and coherent work terms. In this way, the present formalism indicates that quantum coherence does not need to be interpreted as an additional independent energetic quantity beyond heat and work. Instead, coherence modifies the usual thermodynamic contributions through quantum dynamical effects. This provides a clearer interpretation for the role of coherence in the quantum first law and helps address ambiguities present in previous formulations, especially regarding the connection between coherence contributions and entropy-related quantities. The perturbative expansion also reveals different physical regimes. At first order, the perturbation generates quantum superpositions without changing the populations of the energy levels. At second order, transition probabilities start to play an active role in the energy balance, leading to irreversible contributions associated with heat and work. In this regime, the formalism becomes naturally connected to transition rates described by Fermi’s golden rule, establishing a direct relation between microscopic quantum transitions and thermodynamic quantities. The present approach provides a useful framework for the study of coherence-driven thermodynamic processes in externally driven quantum systems. Possible applications include quantum thermal machines, driven qubits, coherent control protocols, and nonequilibrium quantum devices operating beyond standard weak-driving or Markovian conditions. The perturbative structure developed here also opens perspectives for future investigations involving higher-order corrections, stronger driving regimes, and possible experimental signatures associated with coherent heat and coherent work.

Overall, we expect these results to contribute to a better understanding of the energetic role of quantum coherence and to provide useful tools for future developments in quantum thermodynamics, particularly in scenarios where coherent dynamics play an important role.

\appendix

\section{First-order quantum work – details}\label{ap1}

The first-order correction can be written as
\begin{align}\label{w11}
\nonumber W^{(1)}(t) &=W_a^{(1)}(t)+W_b^{(1)}(t)\\\nonumber&=
\int_{0}^{t}\,dt'\sum_{n,k}
      \rho_{k}^{(1)}(t')\,\big|\langle n\,|\,k\rangle\big|^{2}\,
      \frac{dE_{n}(t')}{dt'}'+\\&2
\int_{0}^{t}\,dt' \sum_{n,k}
      \rho_{k}^{(0)}
      \,\mathrm{Re}\left\{\langle k\,|\,n\rangle c^{(1)}_{nk}(t')\right\}
      \frac{dE_{n}(t')}{dt'}
\end{align}
Let us focus on the first term of the above equation
\begin{align}
\nonumber W^{(1)}_a(t)&=
\int_{0}^{t}dt'\sum_{n,k}
      \rho_{k}^{(1)}(t')\,\big|\langle n\,|\,k\rangle\big|^{2}\,
      \frac{dE_{n}(t')}{dt'}
\\\nonumber&=\int_{0}^{t}\,dt'\sum_{k}\rho_{k}^{(1)}(t')\sum_{n}\,\big|\langle n\,|\,k\rangle\big|^{2}\,
      \dot E_{n}(t')
\\&=\int_{0}^{t}\,dt'\sum_{k}\rho_{k}^{(1)}(t')\langle k\,|\dot {\hat H}(t')|\,k\rangle
\end{align}
Here, we have used the Feynman–Hellmann theorem \cite{Reis2025}  $\partial E_n/\partial t=\langle n|\partial H/\partial t|n\rangle$. This first term of the first-order correction to the work can be further developed by considering $\rho_{k}^{(1)}$. Then, we have
\begin{align}
W^{(1)}_a(t)&=
-\frac{i}{\hbar}\,
\int_{0}^{t}dt'\;
      \int_{0}^{t'}d t''\;\sum_k
      \bigl\langle k\bigl|\,
            \bigl[\hat{V}_I(t''),\hat{\rho}^{(0)}\bigr]\dot {\hat H}(t')
      \bigr|k\bigr\rangle
\end{align}

We now consider the second term of Eq.~\eqref{w11}
\begin{align}\label{segere}
W^{(1)}_b(t)=2
\int_{0}^{t}\,dt' \sum_{n,k}
      \rho_{k}^{(0)}
      \,\mathrm{Re}\left\{\langle k\,|\,n\rangle c^{(1)}_{nk}(t')\right\}
      \frac{dE_{n}(t')}{dt'}
\end{align}
We can write
\begin{align}
\nonumber 2\,\text{Re}&\,\left\{\langle k|n\rangle\,c^{(1)}_{nk}(t')\right\}
=\langle k|n\rangle\,c^{(1)}_{nk}(t')+\langle n|k\rangle\,[c^{(1)}_{nk}(t')]^{*}
\\&=\frac{i}{\hbar}\int_{0}^{t}\!dt'
\Big[-\langle k|n\rangle\,\langle n|\hat V_I(t')|k\rangle
+\langle n|k\rangle\,\langle k|\hat V_I(t')|n\rangle\Big].
\end{align}
Considering
\[
-\,\langle k|n\rangle\,\langle n|\hat V_I|k\rangle
+\langle n|k\rangle\,\langle k|\hat V_I|n\rangle
=\langle k|\,[\,\hat V_I,\,|n\rangle\langle n|\,]\,|k\rangle,
\]
we obtain
\begin{equation}\label{plmju}
2\,\text{Re}\left\{\langle k|n\rangle\,c^{(1)}_{nk}(t')\right\}
=\frac{i}{\hbar}\int_{0}^{t'}\!dt''
\langle k|\,[\,\hat V_I(t''),\,|n\rangle\langle n|\,]\,|k\rangle.
\end{equation}
Equation \eqref{segere} can be rewritten as
\begin{align}\label{rrefff}
\nonumber &W_b^{(1)}(t)
=\\& \frac{i}{\hbar} \int_{0}^{t} dt' \int_{0}^{t'} dt'' \sum_{n,k} \; \rho_k^{(0)}\,\dot{E}_n(t')  \;
\langle k | \left[ \hat V_I(t''), \, |n\rangle\langle n| \right] | k \rangle .
\end{align}
Considering the Feynman–Hellmann theorem \cite{Reis2025}  $\partial E_n/\partial t=\langle n|\partial \hat H/\partial t|n\rangle$ e $\rho_k^{(0)} = \langle k|\hat{\rho}^{(0)}|k\rangle$, we can write
\begin{align}\label{rrefff2}
W^{(1)}_b(t)&=-\frac{i}{\hbar}\,
\int_{0}^{t}dt'
      \int_{0}^{t'}d t''\sum_k
      \bigl\langle k\bigl|\,
            \bigl[\hat{V}_I(t''),\hat{\rho}^{(0)}\bigr]\dot {\hat H}(t')
      \bigr|k\bigr\rangle.
\end{align}

Then, we have the following result for the first-order correction of the quantum work
\begin{align}\label{w112}
\nonumber W^{(1)}(t)&=W_a^{(1)}(t)+W_b^{(1)}(t)=\\&-\frac{2i}{\hbar}\,
\int_{0}^{t}dt'
      \int_{0}^{t'}d t''\sum_k
      \bigl\langle k\bigl|\,
            \bigl[\hat{V}_I(t''),\hat{\rho}^{(0)}\bigr]\dot {\hat H}(t')
      \bigr|k\bigr\rangle.
\end{align}

\newpage
\begin{widetext}
\section{Second-order quantum work – details}\label{apena}

We start with the general second-order contribution

\begin{align}
W^{(2)}(t)
&=\int_{0}^{t}\!dt' \sum_{n,k}\Big[
\rho_k^{(2)}(t')\,\big|c^{(0)}_{nk}\big|^{2}
+2\,\rho_k^{(1)}(t')\,\text{Re}\big\{\,[c^{(0)}_{nk}]^{*}\,c^{(1)}_{nk}(t')\big\}
+\rho_k^{(0)}\,\big|c^{(1)}_{nk}(t')\big|^{2}
\Big]\;\dot E_n(t') \nonumber \\[0.4cm]
&= W_a^{(2)}(t)+W_b^{(2)}(t)+W_c^{(2)}(t),
\end{align}
\end{widetext}
where
\begin{align}
W_a^{(2)}(t)
&=\int_{0}^{t}\,dt' \sum_{n,k} \rho_k^{(2)}(t')\,\big|c_{nk}^{(0)}\big|^{2}\,\dot E_n(t'), \\
W_b^{(2)}(t)
&=2\int_{0}^{t}\,dt' \sum_{n,k}\rho_k^{(1)}(t')\,
\text{Re}\left\{\big[c_{nk}^{(0)}\big]^{*}c_{nk}^{(1)}(t')\right\}\,
\dot E_n(t'), \\
W_c^{(2)}(t)
&=\int_{0}^{t}\,dt' \sum_{n,k}\rho_k^{(0)}\,\big|c_{nk}^{(1)}(t')\big|^{2}\,\dot E_n(t').
\end{align}

Let us now examine the first contribution
\[
W_a^{(2)}(t)=\int_{0}^{t}\,dt' \sum_{n,k} \rho_k^{(2)}(t')\,|\langle n|k\rangle|^{2}\,\dot E_n(t'),
\]  
by using the Feynman–Hellmann theorem
\(\dot E_n(t')=\langle n|\dot {\hat H}(t')|n\rangle\). Accordingly, the above expression can be recast as
\[
W_a^{(2)}(t)=\int_{0}^{t}\,dt'\sum_{n,k}\rho_k^{(2)}(t')\,|\langle n|k\rangle|^{2}\,\langle n|\dot {\hat H}(t')|n\rangle.
\]  
Considering
\[
\sum_{n}|\langle n|k\rangle|^{2}\,\langle n|\dot {\hat H}(t')|n\rangle = \langle k|\dot {\hat H}(t')|k\rangle,
\]  
we obtain
\begin{equation}\label{dedededqq}
W_a^{(2)}(t)=\int_{0}^{t}\,dt' \sum_{k}\rho_k^{(2)}(t')\,\langle k|\dot {\hat H}(t')|k\rangle .
\end{equation}
Exploring the term $\rho^{(2)}_{k}$ with respect to time, we obtain
\begin{align}\label{der1}
\frac{d\hat{\rho}_{I}(t)}{dt} &=
      \frac{d\hat{U}_{I}}{dt}\,\hat{\rho}_{I}(0)\,\hat{U}_{I}^{\dagger}(t)
      + \hat{U}_{I}(t)\,\hat{\rho}_{I}(0)\,\frac{d\hat{U}_{I}^{\dagger}}{dt}.
      \end{align}
Using Eq.~\eqref{dwede3221q2q} we arrive at
\begin{align}
\frac{d\hat{\rho}_{I}(t)}{dt} &=
      -\frac{i}{\hbar}\,\bigl[\hat{V}_{I}(t),\,\hat{\rho}_{I}(t)\bigr].
\end{align}
Then, we obtain
\begin{align}\label{r2o}
\nonumber \dot\rho^{(2)}_{k}(t)
&=-\frac{i}{\hbar}\,\langle k|[\hat V_I(t),\hat\rho^{(1)}_{I}(t)]|k\rangle\\& =-\frac{i}{\hbar}\sum_{n}\!\Big(V_{kn}\rho^{(1)}_{nk}-\rho^{(1)}_{kn}V_{nk}\Big).
\end{align}
However
\begin{align}
\rho^{(1)}_{nk}(t)
&=-\frac{i}{\hbar}\int_{0}^{t}\!dt'\,\langle n|[\hat V_I(t'),\hat \rho(0)]|k\rangle \nonumber\\
&=-\frac{i}{\hbar}\int_{0}^{t}\!dt'\,[\rho_k(0)-\rho_n(0)]\,V_{nk}(t')  \nonumber \\
&=\big[\rho^{(0)}_{k}-\rho^{(0)}_{n}\big]\,c^{(1)}_{nk}(t).
\end{align}
Substituting the above result into Eq.~\eqref{r2o} we obtain
\begin{equation}
\dot\rho^{(2)}_{k}(t)=-\frac{1}{\hbar}\sum_{n}
\Big(V_{kn}c^{(1)}_{nk}(t)+c^{(1)}_{kn}(t)V_{nk}\Big)\,
\big[\rho^{(0)}_{k}-\rho^{(0)}_{n}\big].
\end{equation}

To further proceed, we need to examine the equation below
\begin{align}
\frac{d}{dt}\big|c^{(1)}_{nk}(t)\big|^{2}
=\dot c^{(1)}_{nk}(t)\,c^{(1)\!*}_{nk}(t)
+c^{(1)}_{nk}(t)\,\dot c^{(1)\!*}_{nk}(t)
\end{align}
where
\begin{align}
\dot c^{(1)}_{nk}(t)=-\frac{i}{\hbar}V_{nk}(t),
\qquad
\dot c^{(1)\!*}_{nk}(t)=\frac{i}{\hbar}V_{kn}(t),
\end{align}
leading to
\begin{align}
\frac{d}{dt}\big|c^{(1)}_{nk}(t)\big|^{2}
=-\frac{i}{\hbar}\Big(V_{kn}(t)c^{(1)}_{nk}(t)+c^{(1)}_{kn}(t)V_{nk}(t)\Big).
\end{align}
Therefore
\begin{align}
\dot\rho^{(2)}_{k}(t)
&=\sum_{n}\frac{d}{dt}\big|c^{(1)}_{nk}(t)\big|^{2}\,
\big[\rho^{(0)}_{k}-\rho^{(0)}_{n}\big]\\
&=\sum_{n}\rho^{(0)}_{k}\frac{d}{dt}\big|c^{(1)}_{nk}(t)\big|^{2}\,
-\sum_{n}\rho^{(0)}_{n}\frac{d}{dt}\big|c^{(1)}_{kn}(t)\big|^{2},\label{eq2222}
\end{align}
leading to
\begin{align}
\rho^{(2)}_{k}(t)
&=\sum_{n}\rho^{(0)}_{k}\big|c^{(1)}_{nk}(t)\big|^{2}
-\sum_{n}\rho^{(0)}_{n}\big|c^{(1)}_{kn}(t)\big|^{2}.
\end{align}

Thus, Eq.~\eqref{dedededqq} can be written as
\begin{widetext}
\begin{equation}\label{ewwewppa}
W_a^{(2)}(t)=\int_{0}^{t}\,dt'\sum_{k}\langle k|\dot {\hat H}(t')|k\rangle \Bigg[\sum_{n}\rho_k^{(0)}\,|c^{(1)}_{nk}(t')|^{2}-\sum_{n}\rho_n^{(0)}\,|c^{(1)}_{kn}(t')|^{2}\Bigg].
\end{equation}
In the second summation, we can swap the indices \(k \leftrightarrow n\); the two terms then acquire the same structure, allowing them to be combined. The final result can thus be written in a symmetric form as 
\begin{equation}\label{ewwewpp}
W_a^{(2)}(t)=\int_{0}^{t}dt'\,\sum_{n,k}\rho_k^{(0)}\,|c^{(1)}_{nk}(t')|^{2}\,\Big[\langle k|\dot {\hat H}(t')|k\rangle-\langle n|\dot {\hat H}(t')|n\rangle\Big].
\end{equation}  
\end{widetext}
Considering the second term of the second-order work correction, we obtain
\[
W_b^{(2)}(t)
= 2\int_{0}^{t}\,dt' \sum_{n,k}\rho_k^{(1)}(t')\text{Re}\Big\{[c^{(0)}_{nk}]^{*}\,c^{(1)}_{nk}(t')\Big\}\;
\dot E_n(t'),
\]
where $c^{(0)}_{nk}=\langle n|k\rangle$. By substituting Eq.~\eqref{plmju} along with the Feynman–Hellmann theorem, $\dot E_n(t')=\langle n|\dot {\hat H}(t')|n\rangle$ into the equation above, we find
\[
W_b^{(2)}(t)
=\frac{i}{\hbar}\int_{0}^{t}\!dt'\int_{0}^{t'}\!dt''\;\sum_{k}\;\rho_k^{(1)}(t')\;
\langle k|\,[\,\hat V_I(t''),\,\dot {\hat H}(t')\,]\,|k\rangle.
\]
We now express the first-order density correction in the interaction picture,
\begin{equation}\label{rho11b}
\rho_k^{(1)}(t')=\langle k|\hat \rho^{(1)}_{I}(t')|k\rangle
=-\frac{i}{\hbar}\int_{0}^{t'}\!\langle k|\,[\,\hat V_I(t''),\,\hat \rho^{(0)}_{I}\,]\,|k\rangle\,dt''.
\end{equation}
If \(\hat \rho^{(0)}_{I}\) is diagonal in \(\{|k\rangle\}\) basis, then 
\(\langle k|\,[\,\hat V_I(t''),\,\hat \rho^{(0)}_{I}\,]\,|k\rangle=V_{kk}(t'')\rho_k^{(0)}-\rho_k^{(0)}V_{kk}(t'')=0\),
consequently
\begin{equation}\label{rho11}
\rho_k^{(1)}(t')=0\quad\text{for every }k,t',
\end{equation}
and, consequently,
\[
W_b^{(2)}(t)=0\quad\text{(if }\hat\rho^{(0)}_{I}\text{ for diagonal in} \{|k\rangle\}\text{ basis).}
\]
If there are initial coherences (i.e., \(\hat\rho^{(0)}_{I}\) is not diagonal in this basis), \(\rho_k^{(1)}(t')\) may be nonzero, and the closed-form expression above shows how \(W_b^{(2)}\) contributes in such cases.

In order to evaluate the third term of the second-order work correction, we employ
\[
W_c^{(2)}(t)
=\int_{0}^{t}\,dt' \sum_{n,k}\rho_k^{(0)}\;|c^{(1)}_{nk}(t')|^{2}\;\dot E_n(t').
\]
By applying the Feynman–Hellmann theorem, we find:  
\[
W_c^{(2)}(t)
=\int_{0}^{t}\!dt'\;\sum_{n,k}\rho_k^{(0)}\;|c^{(1)}_{nk}(t')|^{2}\;\langle n|\dot {\hat H}(t')|n\rangle.
\]
By summing \(W_a^{(2)}\) and \(W_c^{(2)}\) we obtain the second-order correction to the work (since \(W_b^{(2)}=0\)):
\begin{align}
\nonumber W^{(2)}&=W_a^{(2)}(t)+W_c^{(2)}(t)
\\&=\int_{0}^{t}\!dt'\;\sum_{n,k}\rho_k^{(0)}\;|c^{(1)}_{nk}(t')|^{2}\;\langle k|\dot {\hat H}(t')|k\rangle.
\end{align}


\bibliography{bib}

\end{document}